\documentclass[preprint]{agujournal2019}
\usepackage{amsmath}
\usepackage{amssymb} 
\usepackage{url} 
\usepackage{soul}

\journalname{JGR: Planets}

\begin{document}

\title{Plasma Turbulence in the Lunar Environment Across Solar Wind and Magnetotail Conditions: Observations from Chandrayaan-2 Radio Science experiment}

\authors{
Keshav Aggarwal\affil{1}\thanks{Corresponding author},
R. K. Choudhary\affil{2},
Abhirup Datta\affil{1},
and Anshuman Sharma\affil{3}
}

\affiliation{1}{Department of Astronomy, Astrophysics and Space Engineering (DAASE), Indian Institute of Technology Indore, Indore, Madhya Pradesh 453552, India}

\affiliation{2}{Space Physics Laboratory (SPL), Indian Space Research Organization, Vikram Sarabhai Space Centre, Thiruvananthapuram, Kerala 695022, India}

\affiliation{3}{ISRO Telemetry Tracking and Command Network (ISTRAC), Bengaluru, Karnataka 560058, India}

\correspondingauthor{Keshav Aggarwal}{keshavagg1098@gmail.com}

\begin{keypoints}
\item Two-way radio occultation from Chandrayaan 2 used to measure small-scale plasma density fluctuations near the Moon
\item All observations sample kinetic and dissipation scale turbulence rather than large-scale inertial range behavior
\item Plasma variability near the Moon is controlled by local conditions and not by global geomagnetic activity
\end{keypoints}

\begin{abstract}
The Moon's transit between the solar wind and Earth's magnetotail exposes the near-lunar environment to large and rapid variations in plasma density and flow structure. Two-way coherent S-band radio occultation measurements from Chandrayaan-2 were used to quantify electron-density fluctuations integrated along the Earth-Moon line of sight. Observed frequencies were processed to remove geometric Doppler contributions derived from relativistic light-time modeling. The remaining frequency residuals represent the cumulative effect of plasma irregularities along the ray path. Power spectral densities were computed for 54 intervals from 2022, yielding temporal spectral indices in the range $1.05 \le \alpha \le 2.66$, corresponding to spatial indices $4.05 \le p \le 5.66$ indicating ion-kinetic and dissipation-range scales. Of the 54 intervals, 9 were classified as inside the modeled magnetopause, 17 in the bow shock/magnetosheath, and 28 in the solar wind.  Spectral indices measured inside the magnetopause are marginally higher than those in the bow shock and solar wind, though the difference is not statistically significant given the limited magnetotail sample. No measurable correlation is found between spectral slope and geomagnetic activity, indicating that the observed variability is dominated by local plasma structure rather than inner-magnetospheric conditions.
\end{abstract}

\section*{Plain Language Summary}
As the Moon orbits Earth, it moves through different space plasma environments. For most of the month it is exposed to the solar wind, while for several days it passes through Earth's magnetotail, a region of disturbed plasma on the night side of Earth. Because the Moon lacks a global magnetic field and has only a very thin atmosphere, changes in surrounding plasma directly affect the near-lunar environment. This study uses radio signals exchanged between Earth and the Chandrayaan-2 spacecraft to investigate these plasma conditions. Small changes in charged particle density along the signal path produce slight frequency variations. After removing predictable effects from spacecraft and Earth motion, the remaining fluctuations reflect plasma irregularities near the Moon. We analyze 54 observation periods from 2022 and find that the measurements are sensitive mainly to very small plasma structures, where turbulent energy is dissipated rather than transferred across larger scales. Comparisons between magnetotail and solar wind intervals show only minor differences that are not statistically significant. No clear link is found with geomagnetic activity, indicating that local plasma conditions near the Moon dominate the observed variability.

\section{Introduction}

The Moon resides within a highly variable plasma environment governed by the interaction between the supersonic solar wind and Earth's magnetosphere. Owing to the absence of a global magnetic field and the presence of only a surface-bounded exosphere \cite{Bauer1996, Hodges1975, Stern1999}, charged particles impinge directly upon the lunar surface, generating localized photoelectron sheaths, surface charging, and near-surface plasma layers \cite{Stubbs2011, Sarantos2012, Shen2023}. Measurements from the Apollo-era surface packages \cite{Daily1977, Reasoner1972}, Kaguya \cite{Kurata2005}, ARTEMIS \cite{Halekas2005, Halekas2018}, LADEE \cite{Cook2013}, Chandrayaan-1 \cite{Choudhary2016}, and recent Chandrayaan-2 ionospheric analyses \cite{Tripathi2022a, Tripathi2022b, Tripathi2025} consistently demonstrate that the lunar plasma environment is dynamic, tenuous, and highly responsive to changes in solar illumination and upstream solar wind forcing.

For most of its orbit, the Moon is immersed in the nominal solar wind, a quasi-neutral, magnetized plasma with proton densities of a few cm$^{-3}$ and velocities of several hundred km s$^{-1}$ \cite{Rucinski1996, Stern1999}. Solar-wind interaction produces an extended lunar wake \cite{Reasoner1972}, containing rarefaction zones, reflected ions, ambipolar electric fields, and multi-scale fluctuations \cite{Travinek2005, Wang2011}. Ion and electron distributions within the wake are strongly modulated by IMF orientation, convective electric fields, and upstream turbulence \cite{Cravens1987, Zhang2006}. These processes generate coherent structures, sharp gradients, and intermittent bursts of kinetic activity. Recent ARTEMIS observations further demonstrated that kinetic-scale magnetic turbulence around the Moon is spatially nonuniform, with enhanced fluctuation power near the subsolar region, terminator, and wake boundaries, reflecting the strong influence of solar-wind interaction and local plasma instabilities \cite{Luo201}. These results emphasize that the near-lunar plasma environment contains structured kinetic turbulence even in the absence of a global intrinsic magnetosphere.

For approximately five to six days of the synodic month, the Moon traverses Earth's magnetotail \cite{Sridharan2010}. The lobes are characterized by extremely low plasma densities ($<0.1$ cm$^{-3}$), high magnetic-field strengths, and limited large-scale stirring, resulting in weak or intermittent turbulence \cite{Haaland2009, Antonova2021}. The plasma sheet, in contrast, contains densities of $\sim0.1$-1 cm$^{-3}$ and hosts magnetic reconnection, current-sheet flapping, and strong heating \cite{Halekas2018, Buccino2022, Eshleman1973}. The plasma-sheet flanks and magnetosheath transition regions display mixed characteristics of solar wind and magnetospheric plasmas \cite{Ando2012, Ambili2022}, making these regions particularly valuable for cross-regime turbulence comparisons.

This monthly passage through distinct regimes, namely the solar wind, magnetosheath, plasma sheet, lobes, creates a natural experiment for testing how turbulence responds to changes in plasma, density, magnetic topology, and external drivers. Turbulence in collisionless plasmas mediates the nonlinear transfer of energy from large scales to ion and electron kinetic scales where dissipation occurs \cite{Fjeldbo1971}. Solar-wind turbulence typically exhibits power law spectra with slopes between $-3/2$ and $-5/3$ (1D) and $-7/2$ and $-11/3$ (3D) in the inertial range, steepening to $-2$ to $-3$ (1D) near ion and electron scales \cite{Ando2012,Fjeldbo1971, Aggarwal2025a, Aggarwal2025b, Yadav2025}. Magnetospheric turbulence, especially within the plasma sheet, is characterized by strong intermittency, steep kinetic-range spectra, and reconnection-driven structure \cite{Buccino2022, Halekas2018}. In the magnetotail lobes, limited cascade activity leads to weaker turbulence and flatter spectra.

Radio occultation (RO) offers a remote-sensing technique for quantifying electron-density fluctuations along a line of sight. The foundational theory relates refractive-index variations to induced phase and frequency perturbations in a coherent radio signal \cite{Eshleman1973, Fjeldbo1971}. RO has been widely applied to planetary ionospheres and plasma regimes \cite{Imamura2010, Imamura2012, Hinson2017, Kliore2008, Cook2013, Withers2014, Withers2021}. Two-way coherent RO, used in this work, enhances sensitivity by locking the downlink to a ground-based MASER reference through a regenerative transponder and applying a precise turn-around ratio (TAR) \cite{Krisher1993, Withers2014}. Previous studies have shown that the lunar ionosphere is confined below $\sim$100 km altitude \cite{ Tripathi2022, Tripathi2025}, but its fluctuation spectrum and especially its kinetic-scale turbulence remains poorly characterized. RO Doppler fluctuations provide a direct measurement of the temporal power spectral density (PSD) of frequency perturbations imposed by the plasma. If the temporal PSD follows

\begin{equation}
P(f) \propto f^{-\alpha},
\end{equation}
then $\alpha$ quantifies the temporal turbulence index. Under Taylor's frozen-in hypothesis \cite{Taylor1938}, the density inhomogeneities are assumed to be convected without change, so that temporal and spatial scales are related via $k = 2\pi f / V_\text{eff}$, where $V_\text{eff}$ is the effective convection velocity. In this case the temporal frequency fluctuation spectral index $\alpha$ and the 3D spatial density turbulence power-law index $p$ are related by $p = \alpha + 3$ \cite{Chashei2005, Chashei2007}. Related techniques have been applied to probe solar coronal turbulence using S-band signals from the Mars Orbiter Mission and Akatsuki spacecraft during solar conjunction \cite{Aggarwal2026}. While the underlying physical principle is shared - electron-density fluctuations along the ray path modulate the received Doppler frequency - the present study differs in measurement geometry (Earth-Moon line of sight rather than solar conjunction), plasma regime (near-lunar environment and magnetotail rather than solar corona), and signal processing approach. The mathematical framework used here follows \cite{Efimov2003} and \cite{Chashei2007}, and the specific signal processing methodology is described in \cite{Aggarwal2026}.

Chandrayaan-2's two way RO dataset from 2022 offers the first continuous, high-stability measurement of these fluctuations around the entire lunar orbit. By combining Doppler fluctuation spectra with an empirical magnetopause model \cite{Shue1997}, the present work maps each observational interval to its plasma regime and evaluates how turbulence evolves across solar wind and magnetotail environments. In this study, we extract plasma-induced Doppler frequency residuals from two way RO measurements with high precision; compute PSDs and derive spectral indices $\alpha$ for each interval; examine how turbulence varies across plasma regimes encountered by the Moon; and assess whether geomagnetic activity modulates kinetic-scale turbulence at lunar distance.
The results show that all observed intervals correspond to ion-kinetic and electron-dissipation ranges, that intervals inside the magnetotail exhibit modest steepening relative to solar wind conditions, and that geomagnetic activity exerts negligible influence. This establishes two way RO as a powerful tool for probing turbulence in extremely tenuous planetary plasma environments.

\section{Method}

This section discusses the two way experiment, the data acquisition, and subsequent signal processing applied to the received data to convert two way coherent Doppler measurements from Chandrayaan-2 into quantitative indices of plasma turbulence near the Moon. The methodology consists of three main stages: (1) identifying the geometric and plasma environment associated with each observation, (2) calibrating and conditioning the raw Doppler measurements to remove deterministic and instrumental contributions, and (3) extracting the spectral properties of the remaining frequency fluctuations and relating them to underlying plasma turbulence. Each step is designed to suppress non-plasma signatures while preserving the stochastic fluctuations produced by electron-density irregularities along the Earth-Moon line of sight.

The Chandrayaan-2 radio occultation experiment employs a two way coherent S-band radio link between the spacecraft and a terrestrial ground station, as described in detail by \citeA{Tripathi2022} \& \citeA{Tripathi2025}. In this configuration, a highly stable uplink signal transmitted from Earth is received by the spacecraft, phase-locked by the onboard transponder, and coherently retransmitted back to the ground station. Consequently, the received downlink signal contains the cumulative phase and frequency perturbations acquired during both the uplink and downlink propagation through the intervening plasma. Because the same ray path is traversed twice, two-way coherent measurements are especially sensitive to refractive effects caused by electron-density irregularities, while many non-common noise sources, such as oscillator instabilities, are strongly suppressed.

The Chandrayaan-2 spacecraft operates in a near-circular orbit at approximately 100 km altitude. At this altitude, the signal path between the spacecraft and the Earth ground station necessarily grazes the near-lunar environment at altitudes of order 0-100 km, within the altitude range of the known lunar ionosphere (confined below $\sim$100 km; \cite{Tripathi2025}). The ray path therefore probes the near-surface plasma layer during each observing session. The illumination region (dayside, nightside, terminator) affects the photoelectron density and hence the ionospheric contribution. The uplink signal at 2041.598 MHz is transmitted from the IDSN ground station, received and coherently retransmitted by the spacecraft at a fixed turnaround ratio of 240/221, and recorded in open-loop mode at IDSN at a sampling cadence sufficient to resolve Doppler fluctuations on timescales of a few seconds. Each observing session spans approximately 10-15 minutes of continuous signal acquisition during a ground station contact pass. Further details of the hardware, calibration chain, and data quality assessment are provided in \citeA{Tripathi2025}. As demonstrated by \citeA{Tripathi2025}, the dominant contribution to short-timescale frequency fluctuations in this configuration arises from plasma-induced phase variations along the Earth-Moon line of sight.

\subsection{Data and Signal processing}

All two-way coherent S-band radio occultation measurements acquired by the Chandrayaan-2 orbiter during 2022 were analyzed. In this mode, the uplink signal at 2041.598 MHz is referenced to an ensemble of Cesium and MASER clock sources at the Indian Deep Space Network (IDSN). The signal propagates through Earth's troposphere and ionosphere, traverses the interplanetary plasma, enters the near-lunar environment, and is received by the spacecraft. The onboard transponder coherently retransmits the signal using a fixed turnaround ratio,

\begin{equation}
\text{TAR} = \frac{240}{221}.
\end{equation}

As a result, the downlink inherits the phase stability of the ground-based reference, and onboard oscillator noise is strongly suppressed. Open-loop recordings at IDSN provide high-cadence measurements of the received downlink frequency. These data include contributions from spacecraft and Earth motion, relativistic effects, and refractive delays introduced by all media traversed by the signal. For each observing interval, the minimum altitude of the radio ray path above the lunar surface (the tangent-point altitude) was computed from SPICE ephemerides using the spacecraft and ground-station positions. The illumination condition at the tangent point (dayside, nightside, or terminator) was determined from the solar zenith angle at the tangent-point location and is listed in Tables 1 and 2. The large majority of observations correspond to terminator geometry, consistent with the viewing conditions of two-way radio occultation from a near-limb spacecraft orbit at 100 km altitude.

\begin{figure}
\centering
\includegraphics[width=0.75\linewidth]{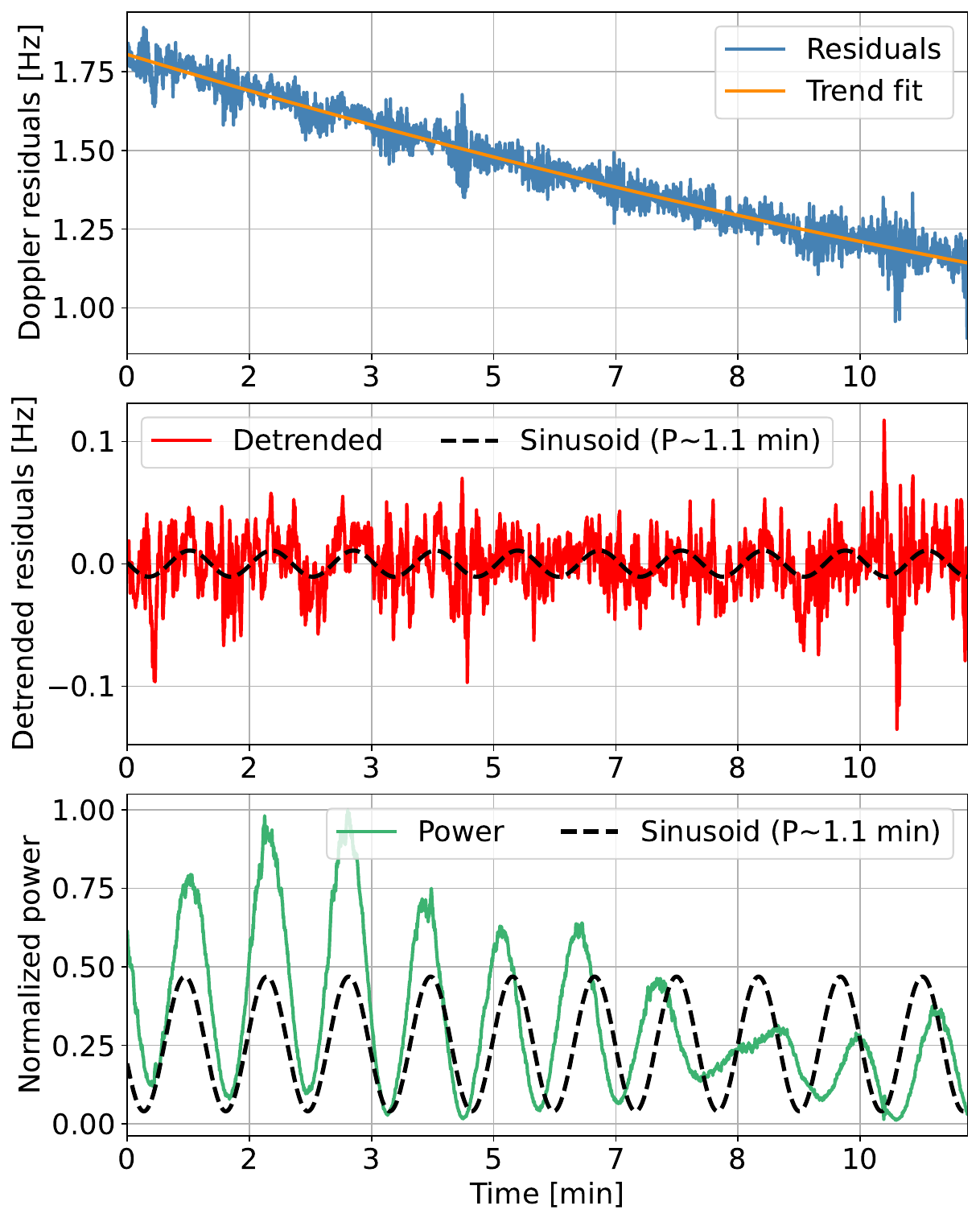}
\caption{Observations from DOY 312. Top panel: Shows time series of Doppler residuals $\Delta f(t)$ (blue) and the second-order polynomial fitted trend used for detrending (orange).
Middle panel: Residuals after polynomial detrending (red) with the observed periodicity indicated (black).
Bottom panel: Power of the received signal (green) with the same periodicity.
Bottom panel: Final detrended residuals after subtraction of the quasi-sinusoidal periodic component. Ultra-low-frequency fluctuations remain.}
\label{fig:signal_processing}
\end{figure}

\begin{figure}
\centering
\includegraphics[width=1\linewidth]{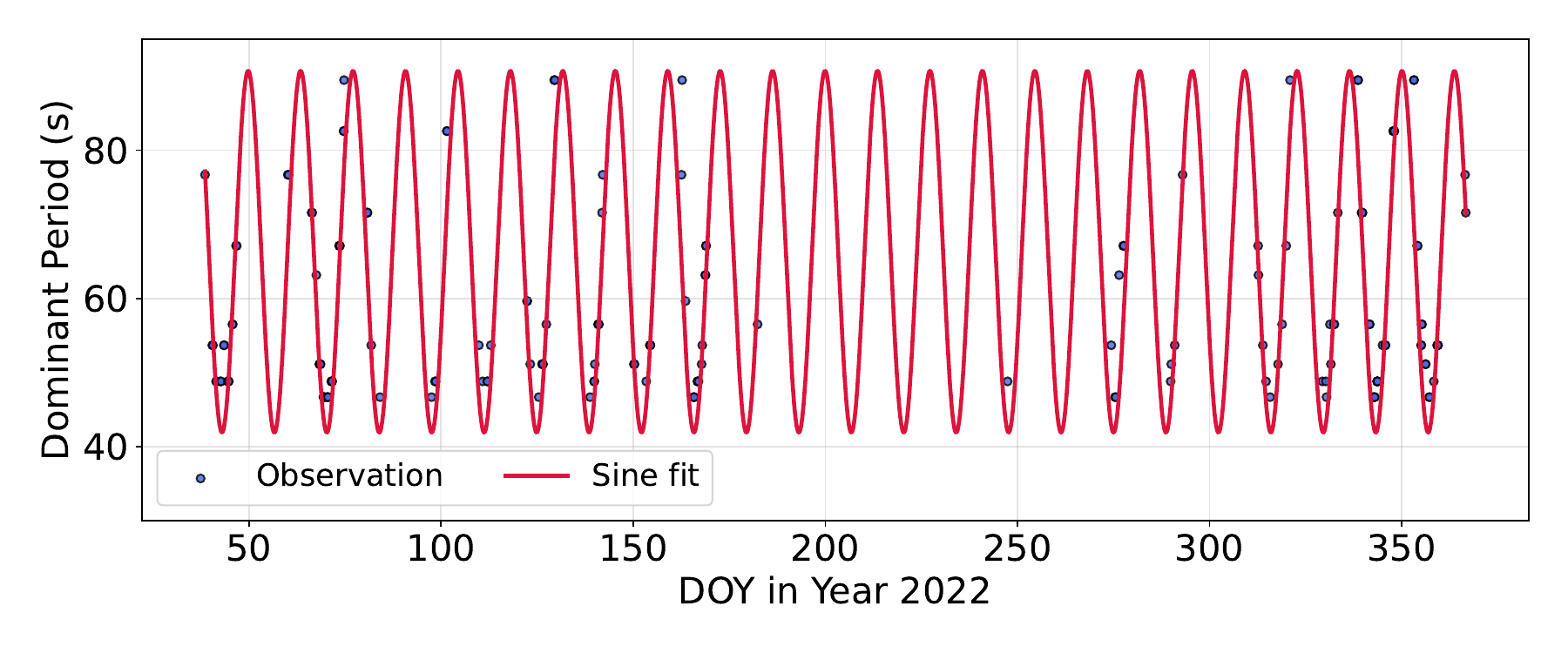}
\caption{Variation of the dominant periodic component extracted from the Doppler residuals as a function of day of year. The points in blue show the period for the day which experiment was conducted, while the curve in red shows the sinusoid fit over these points, yielding a rough time period of $\sim13$ days.}
\label{fig:periodicity}
\end{figure}

The primary observable in this study is the received coherent downlink frequency, $f_{\text{obs}}(t)$. Deterministic Doppler contributions associated with spacecraft motion, Earth rotation, and relativistic light-time effects are removed by computing a predicted frequency, $f_{\mathrm{th}}(t)$, using a relativistic light-time model analogous to that employed by \citeA{Tripathi2025}. The Doppler frequency residuals are defined as

\begin{equation}
\Delta f(t) = f_{\text{obs}}(t) - f_{\mathrm{th}}(t)
\end{equation}
representing cumulative refractive effects of plasma irregularities along the two way propagation path. Values of $\Delta f(t)$ as observed on DOY 312 (08 November 2022) during 18:00 to 18:12 UT are shown in the top panel of Figure \ref{fig:signal_processing}. The relationship between two-way coherent Doppler frequency residuals and electron-density fluctuations is well established in the radio occultation literature \cite{Wexler2019, Jain2022, Jain_2023, Jain_2024}. Doppler frequency residuals $\Delta f$ can be used to estimate the fluctuations in the column electron densities during the period of the measurements as

\begin{equation}
    \Delta N_e = \frac{c f_{Hz}}{\kappa} \Delta f
\end{equation}

where $f_{Hz}$ is the transmitted frequency in Hz, and $\kappa \approx 40.3 m^3/s^2$ \cite{Ando2015, Imamura2005, Miyamoto2014}. Although these residuals contain the desired stochastic plasma fluctuations, they also include smooth, slowly varying trends arising from geometric and instrumental effects. A 2nd order polynomial trend fit (shown by the orange line in this Figure) is applied to remove systematic trend, which is also known as the baseline correction \cite{Tripathi2022b}. The baseline corrected residuals are plotted in the middle panel of Figure \ref{fig:signal_processing} as color red lines. We however note a sinusoidal variation in the Doppler residual (shown in black). A similar behavior is seen in the power of the received signal as shown in the bottom panel of Figure \ref{fig:signal_processing} as green color lines. This behavior is deterministic, arising from spacecraft spin, Chandrayaan-2 orbital motion, and the Moon's orbit around Earth, and is not related to plasma turbulence. Analysis of these detrended residuals, when taken one at a time, shows a periodic component of $\sim$40-120 seconds, while the data when looked altogether shows a quasi-sinusoidal modulation with a characteristic timescale of approximately 13 days (Figure \ref{fig:periodicity}).

The origin of this periodic component is geometric and instrumental rather than plasma-related. It arises from the combined effect of spacecraft attitude motion (including any residual spin or nutation), the periodically varying viewing geometry of the Chandrayaan-2 orbit relative to the IDSN ground station, and the Moon's orbital motion around Earth. The same periodicity is visible in the received signal power (Figure \ref{fig:signal_processing}, bottom panel), confirming its non-plasma origin. The 13-day envelope (Figure \ref{fig:periodicity}) is consistent with the synodic beat between the lunar orbital period and the ground station contact geometry. This deterministic component is modeled and removed prior to spectral analysis following the approach of \cite{Parisi2023}, and the remaining residuals are dominated by stochastic fluctuations suitable for turbulence characterization.
We model this periodic component as

\begin{equation}
r_{\text{sin}}(t) = A_1 \sin(\omega_1 t') + B_1 \cos(\omega_1 t'), \text{where } t' = t - t_0
\end{equation}

\begin{figure}
\centering
\includegraphics[width=1\linewidth]{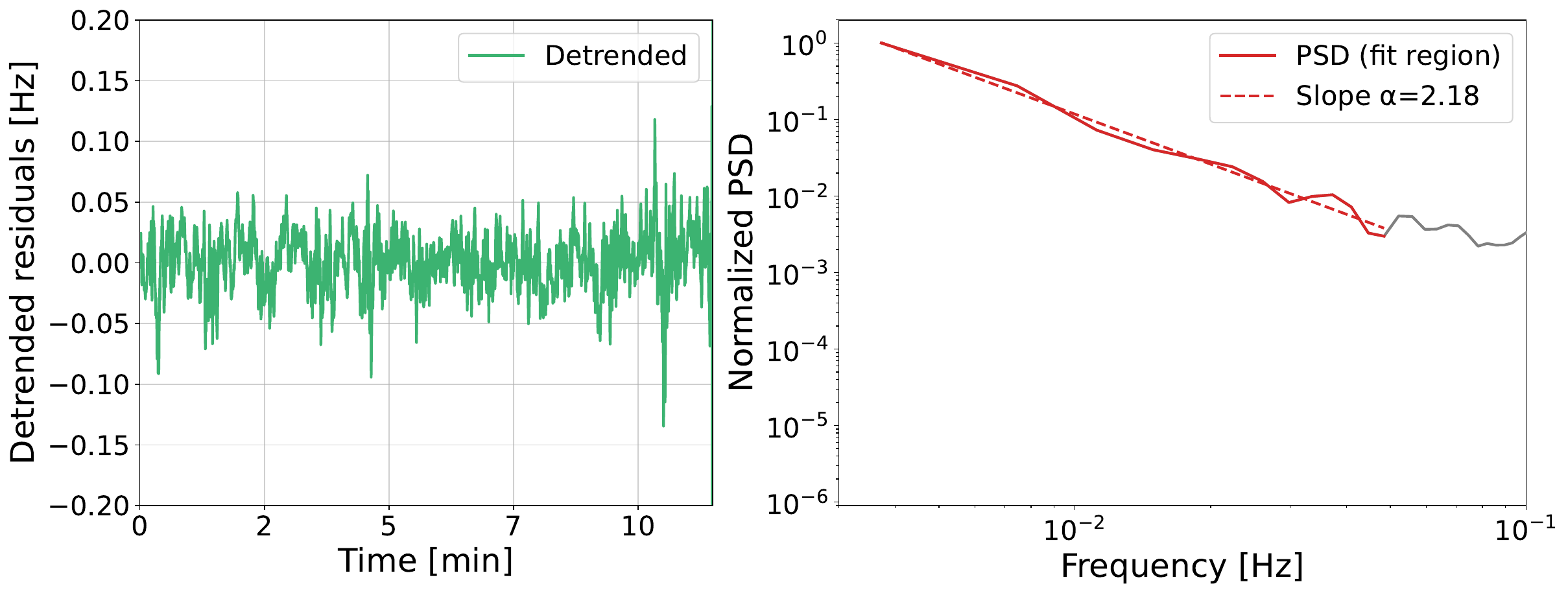}
\caption{Observations from DOY 312. Left panel: Residuals after complete detrending (Doppler frequency fluctuations). Right panel: Normalised Power spectral density (PSD) computed using Welch's method.}
\label{fig:PSD}
\end{figure}
The final detrended residuals serve as input to power spectral density (PSD) estimation. PSD is computed using Welch's method \cite{Welch1967,Yadav2025} and normalized by its maximum value. A power law slope is fitted over the frequency range

\begin{equation}
f_{\min} = \frac{1}{N}, \qquad f_{\max} = \frac{1}{20}\ \text{Hz},
\end{equation}

\noindent where $N$ is the number of samples and $1/20$ Hz corresponds to one-tenth of the Nyquist frequency. PSD values below the noise floor (shown in grey) are excluded from the fit. Figure \ref{fig:PSD} illustrates the final residuals and corresponding PSD for DOY 312.

\section{Results}

A total of 54 Doppler residual intervals from 2022 satisfied all methodological criteria described in Section 2. These intervals sample a wide range of plasma conditions encountered as the Moon traversed the solar wind and the terrestrial magnetotail at different phases of the lunar month.

To characterize the plasma environment associated with each observation, spacecraft ephemerides were obtained from SPICE kernels using the \texttt{spkezr} routine \cite{Acton1996, Annex2020}. The spacecraft position was expressed in a Sun-Earth-aligned coordinate system, where the $+X$ axis points sunward and the $Y$ axis lies perpendicular to the Sun-Earth line in the ecliptic plane. For each epoch, the instantaneous basis vectors were constructed from the Sun-Earth geometry to ensure consistency with the empirical boundary models.

The classification of spacecraft location relative to Earth's magnetosphere was performed using two empirical boundary models evaluated in the same Sun-Earth-aligned geometry: the magnetopause model of \cite{Shue1997},
\begin{equation}
r_{\text{mp}}(\theta) = R_0 \left( \frac{2}{1 + \cos\theta} \right)^{\alpha_{\text{mp}}},
\end{equation}
with $R_0 = 10 R_\oplus$ and $\alpha_{\text{mp}} = 0.6$, and the empirical bow shock model of \cite{Chao2002}, $r_{\text{bs}}(\theta)$, evaluated using the nominal solar wind dynamic pressure and interplanetary magnetic field $B_z$ adopted throughout this study. In both expressions, $\theta$ is the solar zenith angle measured from the Sun-Earth line. For each Doppler interval, the spacecraft radial distance $r_{\text{sc}}$ was compared against both boundaries and classified into one of three plasma regions,
\begin{equation}
r_{\text{sc}} < r_{\text{mp}}(\theta) \Rightarrow \text{magnetopause (inside the magnetotail)},
\end{equation}
\begin{equation}
r_{\text{mp}}(\theta) \le r_{\text{sc}} < r_{\text{bs}}(\theta) \Rightarrow \text{bow shock (magnetosheath)},
\end{equation}
\begin{equation}
r_{\text{sc}} \ge r_{\text{bs}}(\theta) \Rightarrow \text{solar wind}.
\end{equation}
The Chao magnetopause model was additionally evaluated and over-plotted for visualization only, to illustrate the broader solar wind-magnetosphere configuration and to assess the robustness of the region boundaries against model choice; it was not used for the formal classification, which relies on the Shue magnetopause and Chao bow shock boundaries listed above.
As an additional measure of global geomagnetic activity, each observation was paired with the nearest hourly Dst index. This geometric and environmental tagging enables a systematic comparison of turbulence properties across different plasma regimes encountered by the Moon during its monthly orbit. Tables \ref{tab:part1}, \ref{tab:part2}, and \ref{tab:part3} give a summary of the Chandrayaan-2 two-way radio occultation observations acquired during 2022 when the Moon was in the Solar wind, listing the observation date, day of year (DOY), Dst index, illumination geometry at the tangent point (dayside, terminator, or nightside), and the derived temporal ($\alpha$) and spatial ($p$) turbulence spectral indices. Table \ref{tab:part1} list the observations when the Moon was in the Solar wind, table \ref{tab:part2} when the Moon was in the Bow shock, and table \ref{tab:part3} when the Moon was in the Magnetotail region.

\begin{figure}
    \centering
    \includegraphics[width=1.1\linewidth]{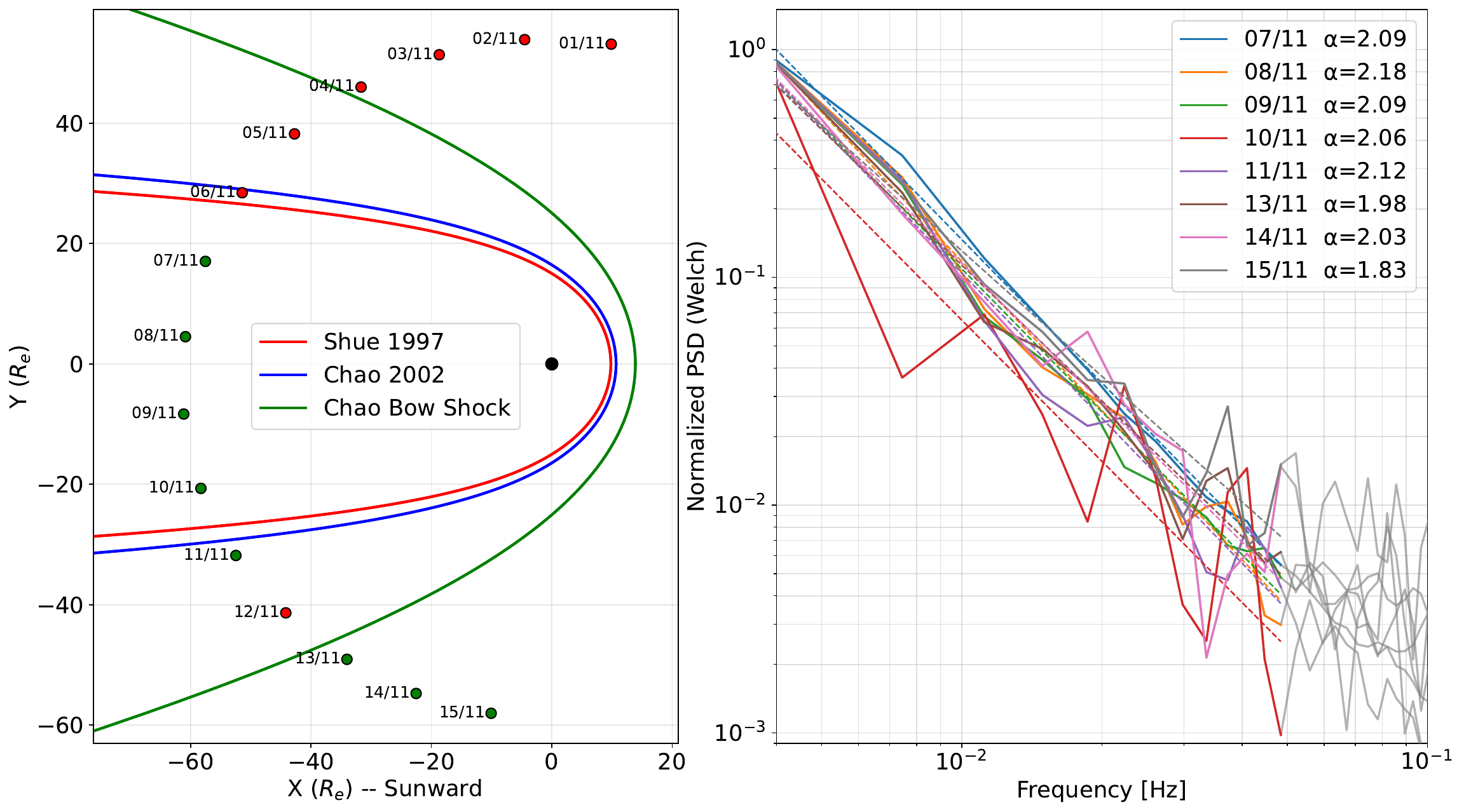}
    \caption{Left panel: This plot shows the movement of the Moon in/out of the magnetotail region defined by \citeA{Shue1997} marked in red, and \citeA{Chao2002} in blue. The curve in green denotes the bowshock as defined by \citeA{Chao2002}. The red and blue points mark the Moon's positions inside/outside the magnetopause, while the green points mark the days Chandrayaan-2 made observations. The black point shows Earth's position. The distances along X and Y axes are shown in respect of $R_e$, the radius of Earth which is equal to 6371 km.  Right: Normalised PSD of observations made by Chandrayaan-2 using the Welch method (unitless). The range considered for the slope fit is shown as a colored solid line, the slope fit is marked with the same color but dashed, and the portion below the noise floor is shown in grey.}
    \label{fig:positions_psd}
\end{figure}

Figure \ref{fig:positions_psd} summarizes the Moon's trajectory relative to the terrestrial magnetosphere together with representative normalised power spectral densities (PSDs) derived from Chandrayaan-2 observations. The left panel shows the Moon's motion with respect to the magnetopause models of \citeA{Shue1997} (red) and \citeA{Chao2002} (blue), and the bow shock model of \citeA{Chao2002} (green). Red and blue symbols indicate the Moon's positions inside and outside the magnetopause, respectively, while the green symbols mark the epochs of Chandrayaan-2 observations. The black symbol denotes the Earth's position. Distances along the X and Y axes are expressed in units of Earth's radius ($R_e = 6371$ km).

The right panel presents the PSDs of representative Chandrayaan-2 observations computed using the Welch method for a subset of intervals spanning DOY 305--319 (1--15 November 2022). This interval was selected because it captures the Moon's transition from the magnetotail (DOY 311--314), through the bow shock/magnetosheath (DOY 315), and into the ambient solar wind (DOY 317--319; Table \ref{tab:part2}), thereby illustrating the spectral variability across the principal plasma regimes examined in this study. The frequency ranges used for spectral fitting are indicated by colored solid lines, the corresponding power-law fits are shown with dashed lines of the same color, and frequency regions below the instrumental noise floor are shaded in grey. The temporal spectral indices, $\alpha$, obtained from the PSD fits span

\begin{equation}
\alpha_{\min} = 1.05, \quad \alpha_{\max} = 2.66,
\end{equation}
with a mean value
\begin{equation}
\overline{\alpha} = 1.82, \quad \sigma_{\alpha} = 0.38,
\end{equation}
and a median of 1.94. All fitted slopes are positive and significantly steeper than would be expected for inertial-range scaling under the adopted definition of the spectral indices. Using the following relationship  

\begin{equation}
p = \alpha + 3,
\end{equation}
each temporal index was converted to its corresponding spatial spectral index. The resulting values lie in the range 

\begin{equation}
4.05 \le p \le 5.66,
\end{equation}
indicating spectra dominated by kinetic-scale and dissipation-range structure. Consequently, none of the measured intervals exhibits inertial-range scaling, and the dataset exclusively probes plasma structure at scales steeper than the inertial cascade. The intervals were further categorized according to the three-region scheme described in Section 2, using the Shue magnetopause and Chao bow shock as the inner and outer boundaries, respectively. Of the 54 intervals, 9 fall inside the magnetopause (magnetotail proper), 17 lie in the intervening bow shock/magnetosheath region, and 28 occur in the upstream solar wind. The corresponding mean spectral indices are
\begin{equation}
\overline{\alpha}_{\text{mp}} \approx 1.91 \; (9 \text{ intervals}), \quad
\overline{\alpha}_{\text{bs}} \approx 1.76 \; (17 \text{ intervals}), \quad
\overline{\alpha}_{\text{sw}} \approx 1.83 \; (28 \text{ intervals}).
\end{equation}
The magnetopause intervals show the steepest mean spectrum, followed by the solar wind and then the bow shock/magnetosheath intervals, but the differences among the three means ($\lesssim 0.15$) remain small compared with the overall dispersion of the dataset ($\sigma_\alpha \approx 0.38$), and no monotonic trend in spectral steepening is observed across the three regions. Given the limited number of intervals in each category, particularly inside the magnetopause, and the substantial overlap between the three distributions, no statistically robust distinction between magnetopause, bow shock, and solar wind slopes can be established.

The Pearson correlation coefficient between $\alpha$ and the Dst index is $r(\alpha,\mathrm{Dst}) \approx -0.09$, with a two-sided p-value of approximately 0.51 for $N=54$, indicating no statistically significant correlation. This is consistent with expectations, as Dst primarily reflects ring-current variations in the inner magnetosphere and is not expected to track small-scale plasma structure at lunar distances, although it provides a useful measure of the overall space weather conditions \cite{Tian2012}.

The PSDs share a common structure: At the lowest frequencies, residual long term trends produce elevated power above the expected scaling. Across the selected fitting interval, the spectra follow an approximate power law, while at high frequencies the power drops sharply as the noise floor is reached. The slopes of the power law region indicate that the near-lunar plasma is dominated by kinetic- and dissipation-range turbulence. The occultation geometry and measurement sensitivity limit access to spatial scales above the ion break, so the inertial range is not observed in these data. A weak dependence on illumination geometry may also be present. The limited number of dayside and nightside observations precludes a robust assessment of illumination effects. While several of the steepest spectra occur during dayside and terminator observations, only three nightside intervals are available, and their spectral indices span a relatively broad range (1.46-1.94). Consequently, any dependence of the spectral slope on illumination geometry should be regarded as tentative. Although the number of nightside observations remains limited, this trend is qualitatively consistent with enhanced plasma structuring near illuminated regions of the lunar surface, where photoelectron production, wake-boundary dynamics, and solar-wind interaction are expected to generate stronger kinetic-scale fluctuations.

\begin{table}
\caption{Summary of the Chandrayaan-2 two-way radio occultation observations acquired during 2022 when the Moon was in the Solar wind, listing the observation date, day of year (DOY), Dst index, illumination geometry at the tangent point (dayside, terminator, or nightside), and the derived temporal ($\alpha$) and spatial ($p$) turbulence spectral indices.}
\label{tab:part1}
\centering
    \begin{tabular}{lllllll}
    \hline
        Date & DOY & Dst  & Location & Illumination & $\alpha$ & p \\ 
          &   & (nT) &   & region &   &   \\ 
          \hline
        9 Feb & 40 & 6 & Solar wind & Terminator & 2.11 & 5.11 \\ 
        10 Feb & 41 & 0 & Solar wind & Terminator & 2.16 & 5.16 \\ 
        11 Feb & 42 & -11 & Solar wind & Terminator & 1.52 & 4.52 \\ 
        12 Feb & 43 & 6 & Solar wind & Terminator & 2.28 & 5.28 \\ 
        11 Mar & 70 & -10 & Solar wind & Terminator & 1.33 & 4.33 \\ 
        12 Mar & 71 & -15 & Solar wind & Terminator & 1.98 & 4.98 \\ 
        25 Mar & 84 & -1 & Solar wind & Terminator & 1.81 & 4.81 \\ 
        11 Apr & 101 & -12 & Solar wind & Terminator & 1.25 & 4.25 \\ 
        12 Apr & 102 & -12 & Solar wind & Terminator & 1.2 & 4.2 \\ 
        22 April & 112 & 1 & Solar wind & Terminator & 2.09 & 5.09 \\ 
        9 May & 129 & -9 & Solar wind & Terminator & 2.11 & 5.11 \\ 
        10 May & 130 & -3 & Solar wind & Dayside & 2.4 & 5.4 \\ 
        20 May & 140 & -8 & Solar wind & Terminator & 2.66 & 5.66 \\ 
        21 May & 141 & -5 & Solar wind & Terminator & 1.6 & 4.6 \\ 
        22 May & 142 & -21 & Solar wind & Terminator & 1.09 & 4.09 \\ 
        23 May & 143 & -9 & Solar wind & Terminator & 1.92 & 4.92 \\ 
        10 Jun & 161 & 4 & Solar wind & Nightside & 1.58 & 4.58 \\ 
        18 Jun & 169 & -5 & Solar wind & Terminator & 2.23 & 5.23 \\ 
        4 Sep & 247 & -54 & Solar wind & Terminator & 2 & 5 \\ 
        2 Oct & 275 & 3 & Solar wind & Terminator & 2.19 & 5.19 \\ 
        3 Oct & 276 & -13 & Solar wind & Terminator & 1.56 & 4.56 \\ 
        4 Oct & 277 & -19 & Solar wind & Terminator & 1.51 & 4.51 \\ 
        16 Oct & 289 & -12 & Solar wind & Terminator & 2.02 & 5.02 \\ 
        13 Nov & 317 & -5 & Solar wind & Terminator & 1.98 & 4.98 \\ 
        14 Nov & 318 & -5 & Solar wind & Terminator & 2.03 & 5.03 \\ 
        15 Nov & 319 & -4 & Solar wind & Terminator & 1.83 & 4.83 \\ 
        13 Dec & 347 & -5 & Solar wind & Terminator & 1.44 & 4.44 \\ 
        14 Dec & 348 & 5 & Solar wind & Terminator & 1.28 & 4.28 \\ 
\hline
\end{tabular}
\end{table}

\begin{table}
\caption{Summary of the Chandrayaan-2 two-way radio occultation observations acquired during 2022 when the Moon was in the Bow shock region, listing the observation date, day of year (DOY), Dst index, illumination geometry at the tangent point (dayside, terminator, or nightside), and the derived temporal ($\alpha$) and spatial ($p$) turbulence spectral indices.}
\label{tab:part2}
\centering
    \begin{tabular}{lllllll}
    \hline
        Date & DOY & Dst  & Location & Illumination & $\alpha$ & p \\ 
          &   & (nT) &   & region &   &   \\ 
          \hline
        13 Feb & 44 & -19 & Bow shock & Terminator & 1.94 & 4.94 \\ 
        14 Feb & 45 & -10 & Bow shock & Terminator & 1.46 & 4.46 \\ 
        14 Mar & 73 & -27 & Bow shock & Terminator & 1.05 & 4.05 \\ 
        15 Mar & 74 & -37 & Bow shock & Terminator & 2.04 & 5.04 \\ 
        21 Mar & 80 & 13 & Bow shock & Terminator & 1.49 & 4.49 \\ 
        22 Mar & 81 & 12 & Bow shock & Terminator & 1.97 & 4.97 \\ 
        19 Apr & 109 & 3 & Bow shock & Terminator & 1.27 & 4.27 \\ 
        20 April & 110 & 3 & Bow shock & Terminator & 1.82 & 4.82 \\ 
        13 May & 133 & -14 & Bow shock & Nightside & 1.46 & 4.46 \\ 
        19 May & 139 & 11 & Bow shock & Dayside & 2.01 & 5.01 \\ 
        11 Jun & 162 & 18 & Bow shock & Terminator & 2.35 & 5.35 \\ 
        12 Jun & 163 & 9 & Bow shock & Terminator & 1.88 & 4.88 \\ 
        6 Oct & 279 & -27 & Bow shock & Terminator & 1.29 & 4.29 \\ 
        11 Nov & 315 & -2 & Bow shock & Terminator & 2.12 & 5.12 \\ 
        4 Dec & 338 & -3 & Bow shock & Nightside & 1.94 & 4.94 \\ 
        5 Dec & 339 & -36 & Bow shock & Terminator & 1.7 & 4.7 \\ 
        11 Dec & 345 & -18 & Bow shock & Terminator & 2.16 & 5.16 \\ 
\hline
\end{tabular}
\end{table}

\begin{table}
\caption{Summary of the Chandrayaan-2 two-way radio occultation observations acquired during 2022 when the Moon was in the magnetotail, listing the observation date, day of year (DOY), Dst index, illumination geometry at the tangent point (dayside, terminator, or nightside), and the derived temporal ($\alpha$) and spatial ($p$) turbulence spectral indices.}
\label{tab:part3}
\centering
    \begin{tabular}{lllllll}
    \hline
        Date & DOY & Dst  & Location & Illumination & $\alpha$ & p \\ 
          &   & (nT) &   & region &   &   \\ 
          \hline
        16 Feb & 47 & -5 & Magnetotail & Terminator & 1.55 & 4.55 \\ 
        16 Mar & 75 & -35 & Magnetotail & Dayside & 2.06 & 5.06 \\ 
        15 May & 135 & -3 & Magnetotail & Terminator & 1.09 & 4.09 \\ 
        18 May & 138 & 11 & Magnetotail & Dayside & 1.72 & 4.72 \\ 
        7 Nov & 311 & -72 & Magnetotail & Terminator & 2.09 & 5.09 \\ 
        8 Nov & 312 & -64 & Magnetotail & Terminator & 2.18 & 5.18 \\ 
        9 Nov & 313 & -42 & Magnetotail & Terminator & 2.09 & 5.09 \\ 
        10 Nov & 314 & -20 & Magnetotail & Terminator & 2.06 & 5.06 \\ 
        7 Dec & 341 & -62 & Magnetotail & Terminator & 2.36 & 5.36 \\ 
\hline
\end{tabular}
\end{table}
\section{Discussion}

Two-way S-band radio occultation measurements from the Chandrayaan-2 orbiter provide a sensitive and reliable method for probing plasma turbulence along the Earth-Moon line of sight. In this technique, the radio signal transmitted between the orbiter and Earth passes through the near-lunar plasma, and its frequency fluctuations carry information about density irregularities and turbulence. After removing deterministic Doppler contributions caused by spacecraft motion and calibrating the residuals, we identified 54 time intervals in 2022 with data suitable for quantitative spectral analysis.

Analysis of these intervals reveals that the temporal power spectral densities (PSDs) follow power law behavior over the observed frequency range. The measured temporal spectral indices $\alpha$ range from 1.05 to 2.66. As $p = \alpha + 3$, which converts temporal to spatial spectra assuming Taylor's frozen-in flow hypothesis, these correspond to spatial indices $4.05 \le p \le 5.66$ \cite{Armand2003}. Such steep indices indicate that all measurements sample the kinetic and dissipation ranges of the turbulent cascade, where energy is transferred to small scales and ultimately dissipated, rather than the larger-scale inertial range. Typical ion kinetic scales in the near-Earth solar wind correspond to proton inertial lengths and gyroradii of order tens to hundreds of kilometers. Under Taylor's frozen-in flow hypothesis, such scales correspond approximately to the sub-Hz to few-Hz: frequency range sampled in the present analysis. Although the line-of-sight integrated nature of the radio occultation measurements prevents identification of a unique ion-scale spectral break, the observed steep spectra are broadly consistent with fluctuations occurring at ion-kinetic and dissipation scales.

To better understand the spectral characteristics of the residual Doppler fluctuations observed in the Chandrayaan-2 datasets, we generated a reference model using pure white Gaussian noise. This ensures that the observed spectral signatures are not misidentified as pure white noise, as stochastic noise signals can also exhibit power-law spectral behavior. In our analysis, we created a normalized white noise time series, scaled it appropriately, and computed its Power Spectral Density (PSD) using Welch's method. This establishes a baseline for comparison with the observed Doppler residuals. Unlike colored noise models, white noise has a flat PSD, which serves as a critical reference to identify departures from purely random fluctuations in the measured data.

\begin{figure}[h!]
\centering
\includegraphics[width=0.95\linewidth]{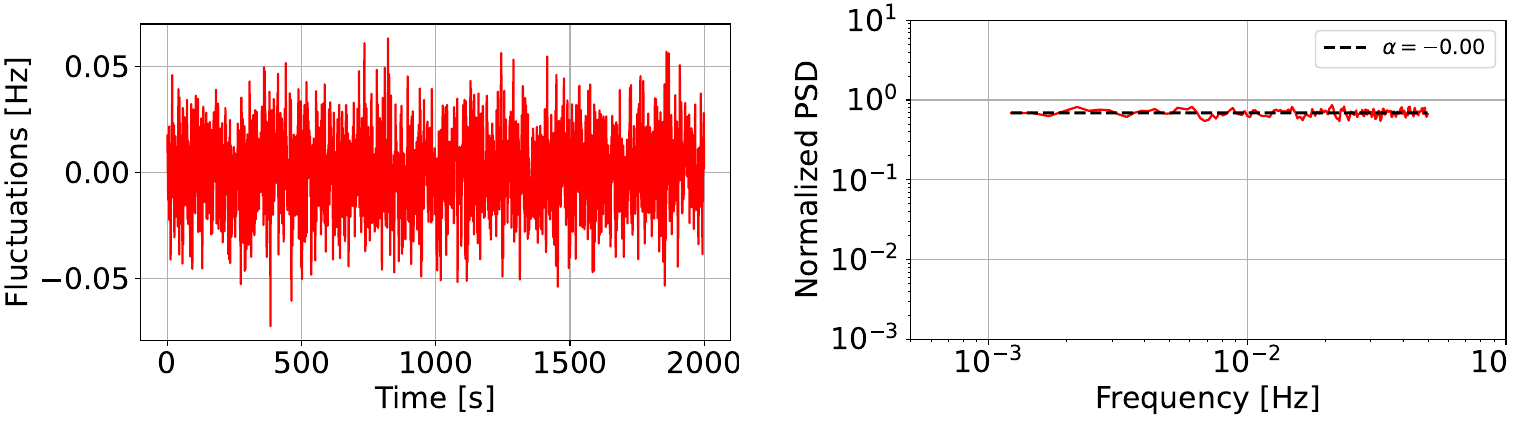}
\caption{Left Panel: Pure white noise generated using Python is shown in red. Right panel: Power Spectral Density of pure white noise generated for comparison with RO Doppler frequency residuals. The PSD is normalized and plotted over a selected frequency range to highlight its flat characteristic.}
\label{fig:psd_noise}
\end{figure}

\vspace*{.25cm}

\noindent The left panel in Figure \ref{fig:psd_noise} shows pure white noise generated using Python, with the amplitude of a similar order as seen in our signals in the IPM, is shown in red. The right panel shows the PSD of this noise for comparison with RO Doppler frequency residuals. The PSD of the generated white noise is normalized and plotted over the selected frequency range to highlight its flat characteristic. Comparing this with the PSDs of the RO Doppler residuals reveals that the observed fluctuations are not flat in frequency but exhibit frequency-dependent behavior. The deviation from the flat PSD of white noise indicates the presence of long-range correlations or structured variability in the plasma along the signal path.

Further, using the three-region classification presented in Section 3 (magnetopause/magnetotail, bow shock/magnetosheath, and solar wind), rather than a simple inside/outside split, we assess whether the plasma regimes intermediate between the solar wind and the magnetotail proper show distinct turbulence properties. Of the 54 intervals, 9 fall inside the magnetopause, 17 lie in the bow shock/magnetosheath, and 28 occur in the solar wind, with mean spectral indices of $\overline{\alpha}_{\text{mp}} \approx 1.91$, $\overline{\alpha}_{\text{bs}} \approx 1.76$, and $\overline{\alpha}_{\text{sw}} \approx 1.83$, respectively. The magnetopause intervals show the steepest mean spectrum, but the pairwise differences among the three regions ($\lesssim 0.15$) remain small compared with the overall dispersion of the dataset ($\sigma_\alpha \approx 0.38$), and no monotonic trend in spectral steepening is observed across regions. Given the small number of intervals sampled inside the magnetopause (9 of 54, $\sim$17\%), this three-way split reduces the statistics available in each bin, and no statistically robust distinction between magnetopause, bow shock, and solar wind turbulence can be established from the present dataset. This imbalance partly reflects the geometry of the lunar orbit: the Moon spends only about five to six days of each synodic month within the magnetotail \cite{Sridharan2010}, compared with roughly three weeks in the solar wind, so that observing sessions distributed across the year inevitably sample the magnetotail less often. The overlap between all three distributions indicates that the kinetic-scale spectral properties are broadly similar across the sampled plasma environments, with at most modest environmental modulation.

The extreme values of the measured slopes illustrate the natural variability of near-lunar plasma. The steepest slope ($\alpha = 2.66$, DOY 140) occurs in the solar wind, and the flattest slope ($\alpha = 1.05$, DOY 73) occurs in the bow shock/magnetosheath; both therefore lie outside the modeled magnetopause.  Without formal uncertainty estimates for the fits, detailed physical interpretation is limited, but these values are consistent with variations in kinetic-scale turbulence rather than transitions between distinct turbulent regimes. Some evidence for spatial or illumination-dependent variability is nevertheless present. Only three nightside intervals are available in the present dataset, with spectral indices ranging from 1.46 to 1.94. Although these values are generally not among the steepest observed, the sample is too small to infer any systematic dependence on illumination geometry. This behavior is qualitatively consistent with previous observations showing enhanced kinetic-scale fluctuations near illuminated regions, wake boundaries, and solar-wind interaction regions around the Moon. However, the present dataset is not sufficiently balanced across viewing geometries to establish this trend statistically.

Each PSD shows an approximately linear region on a log-log plot over the frequency interval used for fitting, justifying the power law assumption. Low-frequency points outside this interval are affected by residual long term Doppler trends and are excluded from the analysis, while high-frequency points drop sharply as the signal approaches the instrument's noise floor. The distribution of fitted slopes consistently indicates that near-lunar plasma turbulence exists primarily in the kinetic and dissipation ranges. While our measurements do not access the inertial range, they provide robust constraints on fine-scale plasma dynamics in a weakly magnetized, low-density environment.

The integrated nature of the line-of-sight measurement means that both the near-lunar ionospheric plasma (at altitudes below $\sim$100 km; \citeA{Tripathi2025}) and the interplanetary plasma column between the Moon and Earth contribute to the observed Doppler fluctuations. At the S-band carrier frequency used here, the doppler frequency residuals are directly related to electron-density fluctuations, and the near-lunar ionospheric contribution, while confined to a thin layer, may contribute a significant fraction of the total Doppler variance during terminator and dayside passes when the photoelectron density is highest. Disentangling the near-lunar ionospheric and interplanetary contributions requires an independent measurement of the ionospheric state, which is outside the scope of the present study, but represents an important direction for future work.

Comparison with other heliospheric environments reinforces the interpretation of our results. In the pristine solar wind near Earth, spacecraft observations show inertial-range scaling at fluid scales, followed by a spectral break near ion kinetic scales and a steeper spectrum at sub-ion scales \cite{Sahraoui2013, Lotz2023, Roberts2023}. High-resolution Parker Solar Probe measurements indicate sub-ion spectral indices between -3 and -5.7, consistent with steep kinetic-range behavior \cite{Lotz2023}. Similar trends are observed in the Saturnian magnetosheath, including regions influenced by Titan, where magnetic fluctuations steepen at small scales, reflecting kinetic-dominated turbulence \cite{Hadid2015}. This widespread occurrence of kinetic-range turbulence supports the conclusion that the near-lunar plasma, as sampled by two way coherent radio occultation, resides in a kinetic and dissipation-dominated regime \cite{Goldstein2015,Howes2008}. The variability in spectral indices likely reflects local plasma structure and wave-mode activity, consistent with observations in interplanetary and planetary magnetospheric turbulence studies.

The physical origin of the observed fluctuations cannot be uniquely identified from the present line-of-sight integrated measurements. In solar wind intervals, the most likely drivers are the ongoing nonlinear turbulent cascade of solar wind fluctuations at these heliocentric distances, supplemented by local processes including reflected-ion beam instabilities in the lunar wake and plasma structuring near the terminator due to differential solar illumination. In magnetotail intervals, relevant mechanisms include current-sheet flapping and reconnection-driven plasma injection in the plasma sheet, as well as shear-flow-driven instabilities at the lobe-to-plasma-sheet boundary. The similarity of spectral indices across regimes at the kinetic scales sampled here is broadly consistent with the expectation that kinetic-scale turbulence develops quasi-universally, while still permitting modest modulation by local plasma conditions and illumination geometry.

\section{Conclusions}
The principal findings of this analysis of 54 two-way coherent S-band radio occultation intervals from Chandrayaan-2 in 2022 can be summarized as follows:
\begin{itemize}
\item All 54 intervals show temporal spectral indices $1.05 \le \alpha \le 2.66$ (spatial $4.05 \le p \le 5.66$), indicating that the measurements are sensitive exclusively to ion-kinetic and dissipation-range turbulence; the inertial range is not accessible with the present occultation geometry.
\item Dividing the intervals into magnetopause (9), bow shock/magnetosheath (17), and solar wind (28) regions gives mean spectral indices of 1.91, 1.76, and 1.83, respectively. The magnetopause mean is marginally the steepest, but the differences between regions are small compared with the overall spread of the data ($\sigma_\alpha \approx 0.38$) and are not statistically robust, given that only 9 of the 54 intervals ($\sim$17\%) sample the magnetopause directly.
\item No significant correlation is found between spectral index and geomagnetic activity (Pearson $r \approx -0.09$ between $\alpha$ and Dst, $N=54$), indicating that global geomagnetic activity does not measurably control the kinetic-scale turbulence sampled near the Moon.
\item Taken together, these results indicate that near-lunar kinetic-scale turbulence is governed primarily by local plasma conditions (the solar-wind cascade, wake and terminator processes, and, within the tail, current-sheet and shear-flow-driven activity) rather than by the large-scale plasma regime or geomagnetic state.
\end{itemize}
We emphasize two limitations of the present analysis. First, the small and uneven sample of magnetotail intervals (9 of 54) is a direct consequence of lunar orbital geometry: the Moon spends only five to six days per synodic month within the magnetotail, compared with roughly three weeks in the solar wind \cite{Sridharan2010}, so the regional comparisons reported here should be regarded as indicative rather than conclusive. Second, the line-of-sight-integrated nature of the measurement mixes contributions from the near-lunar ionosphere and the interplanetary plasma column, which cannot be separated without an independent measurement of the ionospheric state. Addressing both limitations (through observing campaigns timed specifically to lunar magnetotail transits, and through joint analysis with independent ionospheric measurements) is a natural direction for future two-way RO studies with Chandrayaan-2 and subsequent missions. Notwithstanding these limitations, the present results establish two-way coherent radio occultation as a reliable technique for characterizing kinetic-scale plasma turbulence in the tenuous, weakly magnetized near-lunar environment, complementing turbulence studies across the broader heliosphere.

\acknowledgments
 K.A. sincerely thanks Head, HRDD, VSSC, ISRO and Director, SPL for the opportunity to visit SPL as part of this project. The authors greatly appreciate the help from Soumyaneal Banerjee from InSWIM Lab at SPL, and Dr. Keshav R. Tripathi at the University of Tokyo. Help provided by R. S. Simi in providing the data is also gratefully acknowledged. K.A. has received a research fellowship (PMRF-2103356) from the Prime Minister's Research Fellowship (PMRF) scheme, Ministry of Education, Government of India. The authors KA and AD acknowledge the use of facilities procured through the funding via the Department of Science and Technology, Government of India sponsored DST-FIST grant no. SR/FST/PSII/2021/162 (C) awarded to the DAASE, IIT Indore. We express our appreciation and gratitude to all the members of the IDSN-18 and IDSN-32 ground station team, the CH2 operation team, the Flight Dynamics team, the Payload Planning team, the Mission Director, and the data center team for their proactive support in conducting RO experiments in two-way mode. A special thanks to Himanshu Pandey and the ISSDC team for their kind assistance with data dissemination.

\section*{Open Research Section}
Lunar Occultation data from Chandrayaan-2, which can be obtained from the Indian Space Science Data Center (ISSDC) at {\it https://pradan.issdc.gov.in/ch2/} was used for the experiment. Position maps were created using data from the Indian Space Science Data Center (ISSDC) and the NASA SPICE toolkit \cite{Acton1996}. The derived results, and the codes have also been added to a Zenodo archive \citeA{Aggarwal2026c}.
\appendix

\section{PSDs for the other days of the experiment part of this study}
The left panel of these plots shows the movement of the Moon In/out of the magnetotail region defined by \citeA{Shue1997} marked in red, and \citeA{Chao2002} in blue. The curve in green denotes the bowshock as defined by \citeA{Chao2002}. The points in red and blue mark the positions of the Moon inside/outside the magnetopause, while the points in green mark the days for which Chandrayaan-2 made observations. In the right panels, PSDs of the observations done from Chandrayaan-2 using the Welch method are shown. The range considered for the slope fit is shown in colored solid line, and the slope fit is marked in the same color but dashed, while the portion below the noise floor has been shown in grey.

\begin{figure}
    \centering
    \includegraphics[width=1.1\linewidth]{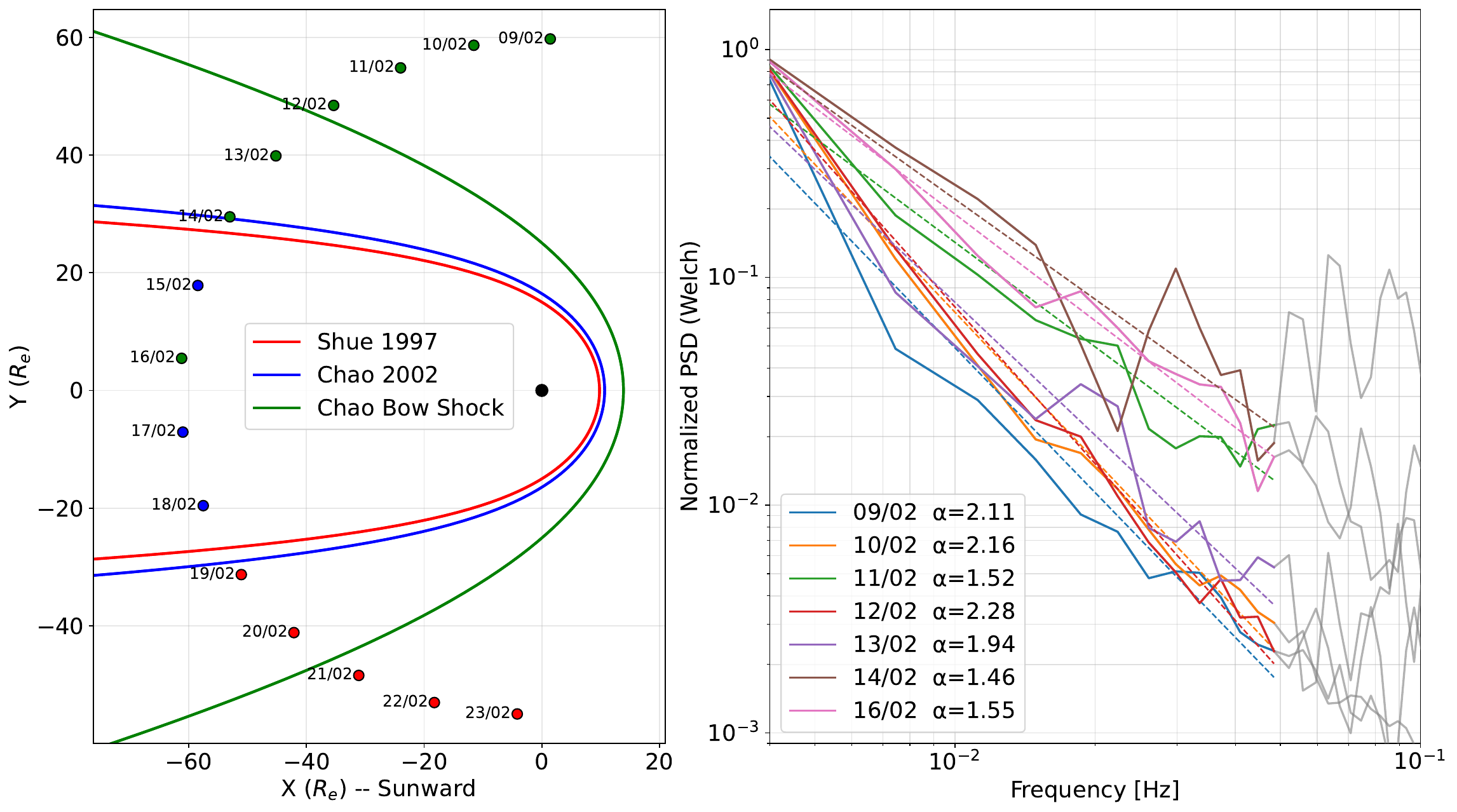}
    \caption{Same as figure \ref{fig:positions_psd} but for DOY 40-54.}
    \label{fig:positions_psd1}
\end{figure}

\begin{figure}
    \centering
    \includegraphics[width=1.1\linewidth]{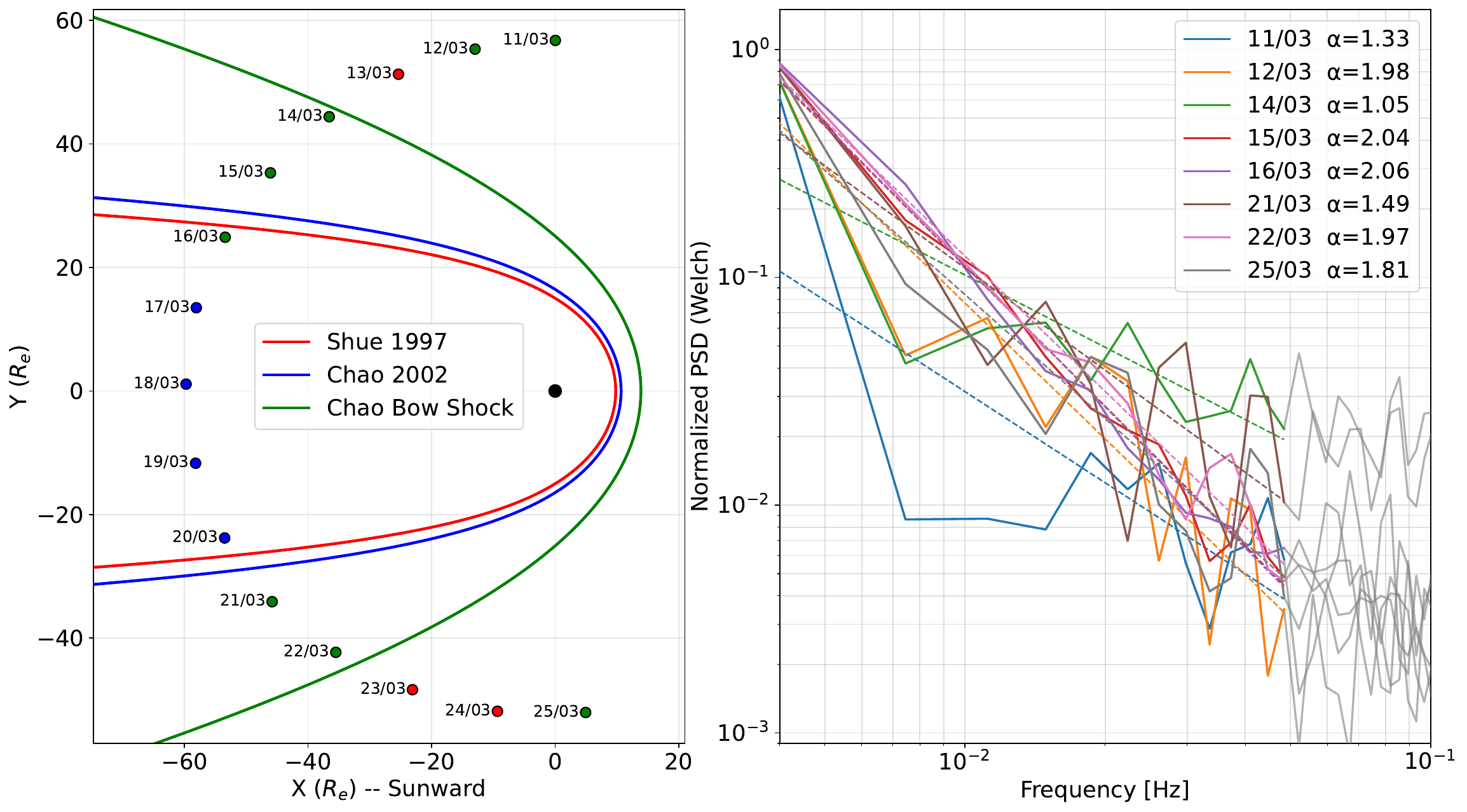}
    \caption{Same as figure \ref{fig:positions_psd} but for DOY 70-84.}
    \label{fig:positions_psd2}
\end{figure}
\begin{figure}
    \centering
    \includegraphics[width=1.1\linewidth]{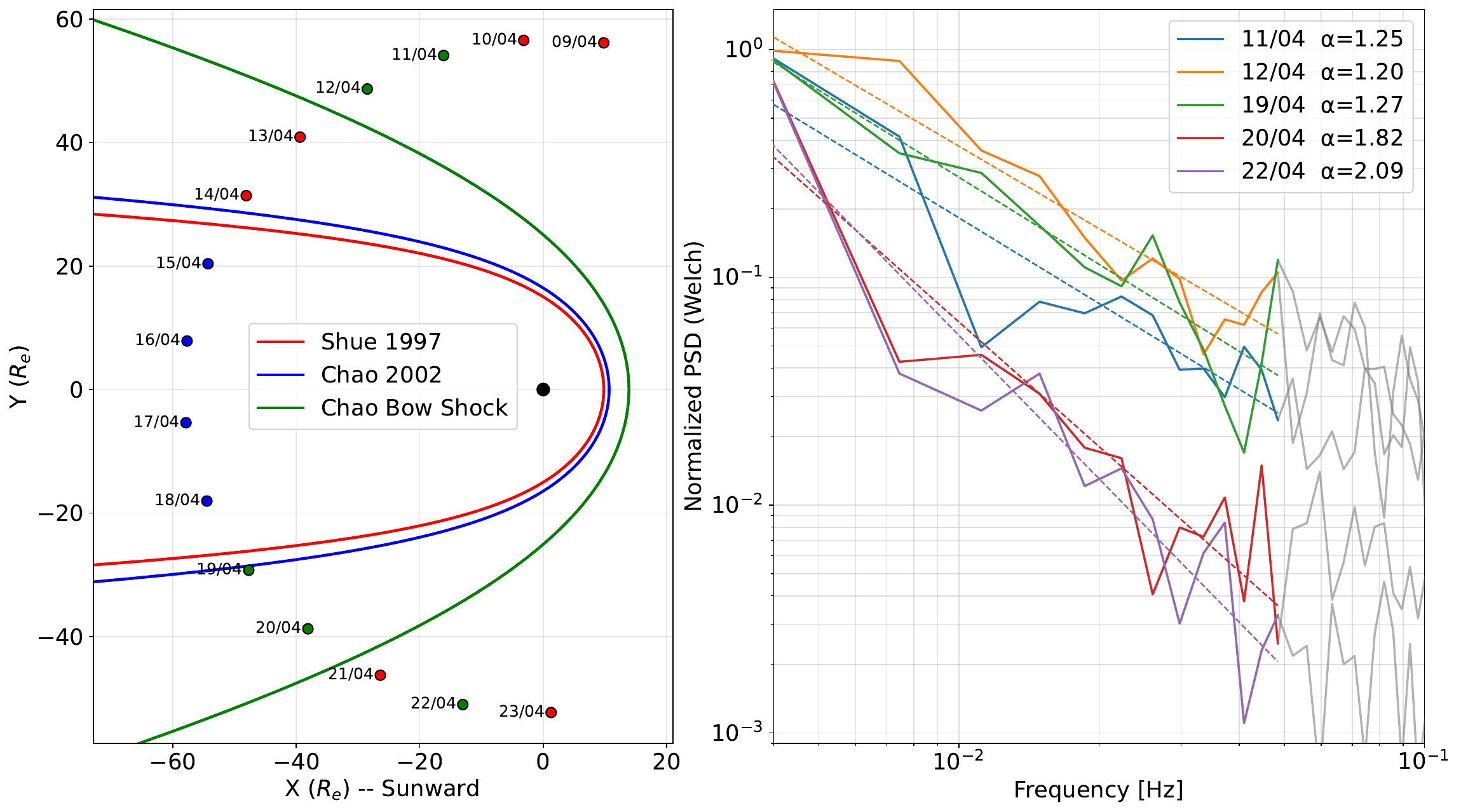}
    \caption{Same as figure \ref{fig:positions_psd} but for DOY 99-113.}
    \label{fig:positions_psd3}
\end{figure}
\begin{figure}
    \centering
    \includegraphics[width=1.1\linewidth]{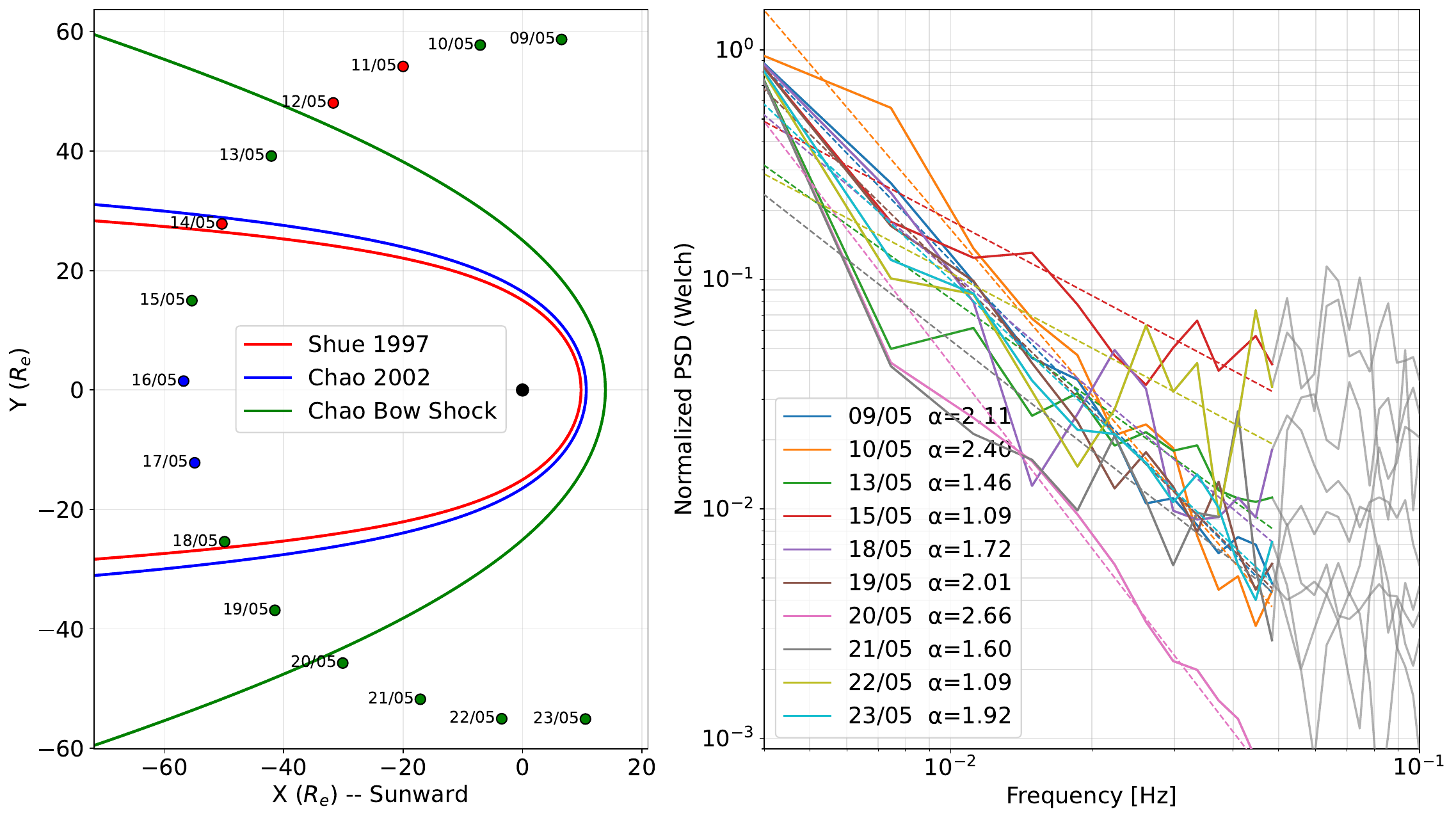}
    \caption{Same as figure \ref{fig:positions_psd} but for DOY 129-143.}
    \label{fig:positions_psd4}
\end{figure}
\begin{figure}
    \centering
    \includegraphics[width=1.1\linewidth]{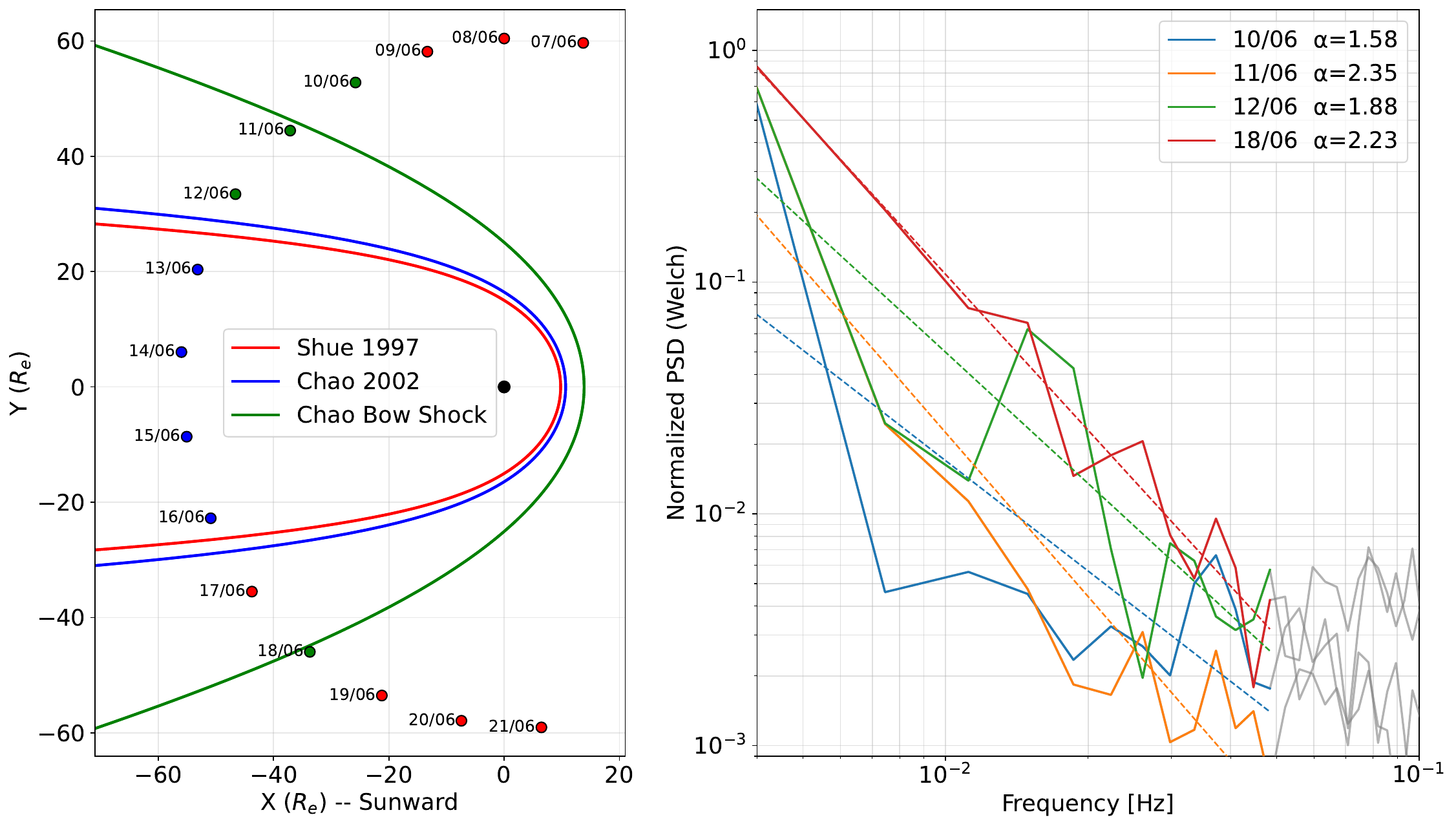}
    \caption{Same as figure \ref{fig:positions_psd} but for DOY 158-172.}
    \label{fig:positions_psd5}
\end{figure}
\begin{figure}
    \centering
    \includegraphics[width=1.1\linewidth]{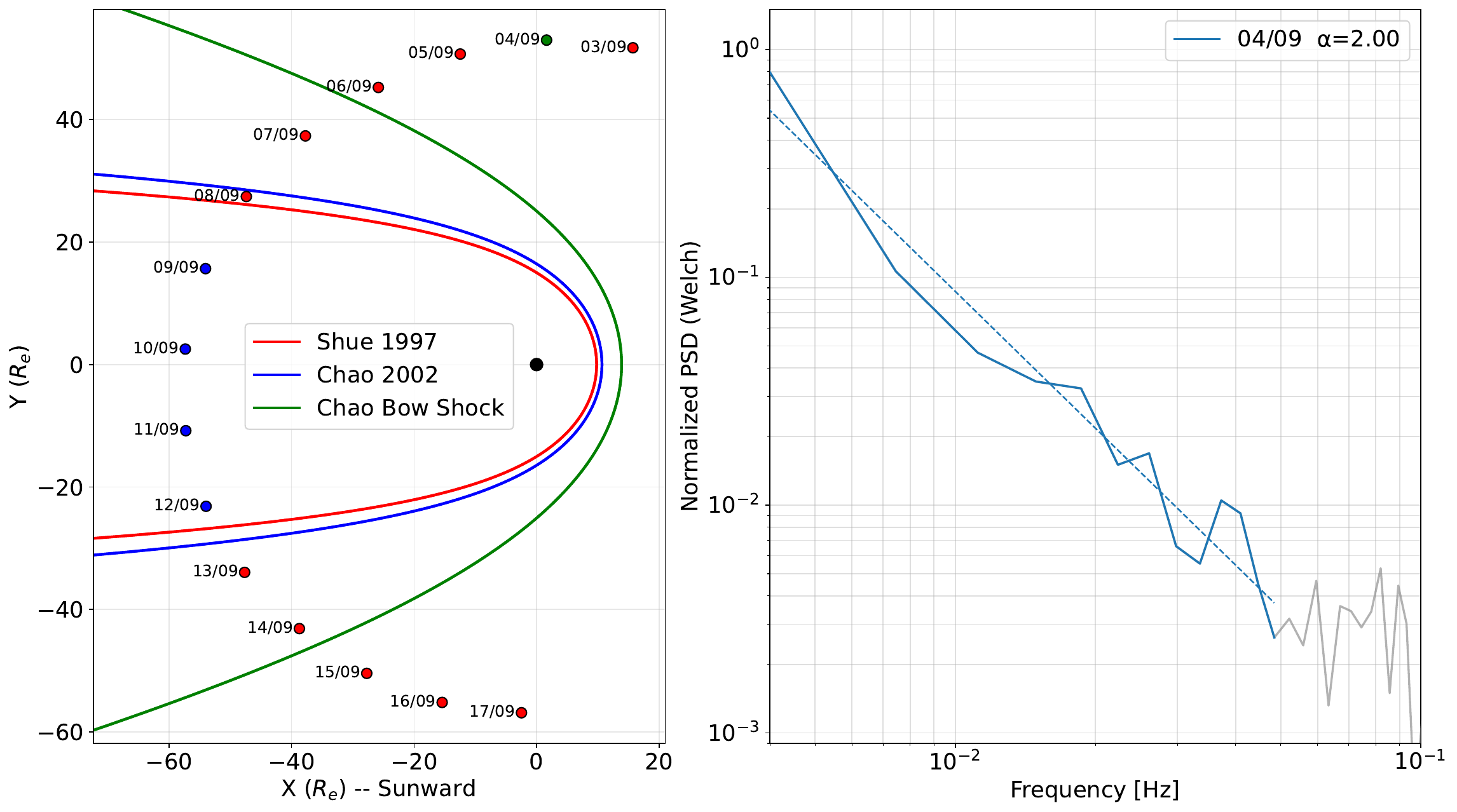}
    \caption{Same as figure \ref{fig:positions_psd} but for DOY 246-260.}
    \label{fig:positions_psd6}
\end{figure}
\begin{figure}
    \centering
    \includegraphics[width=1.1\linewidth]{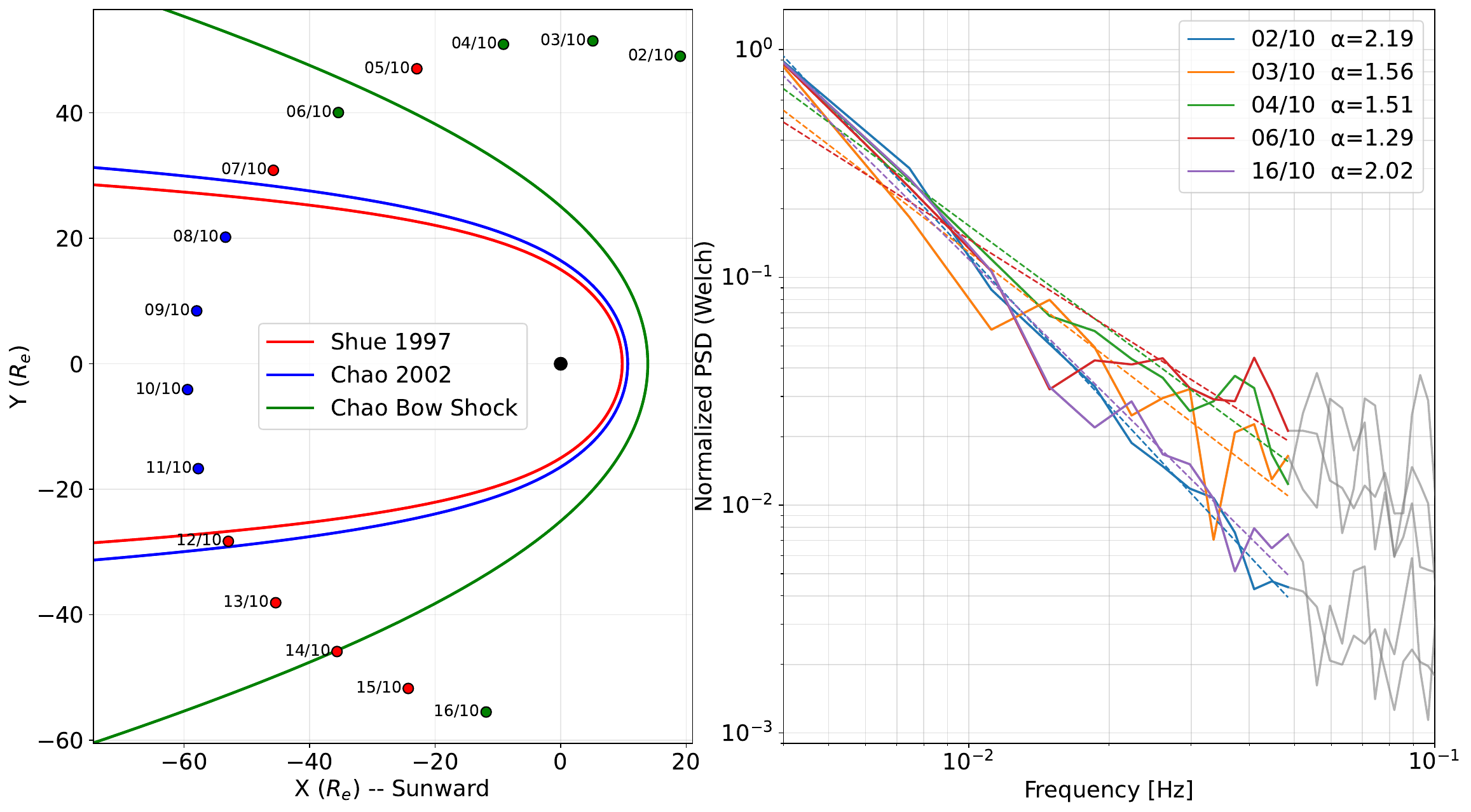}
    \caption{Same as figure \ref{fig:positions_psd} but for DOY 275-289.}
    \label{fig:positions_psd7}
\end{figure}
\begin{figure}
    \centering
    \includegraphics[width=1.1\linewidth]{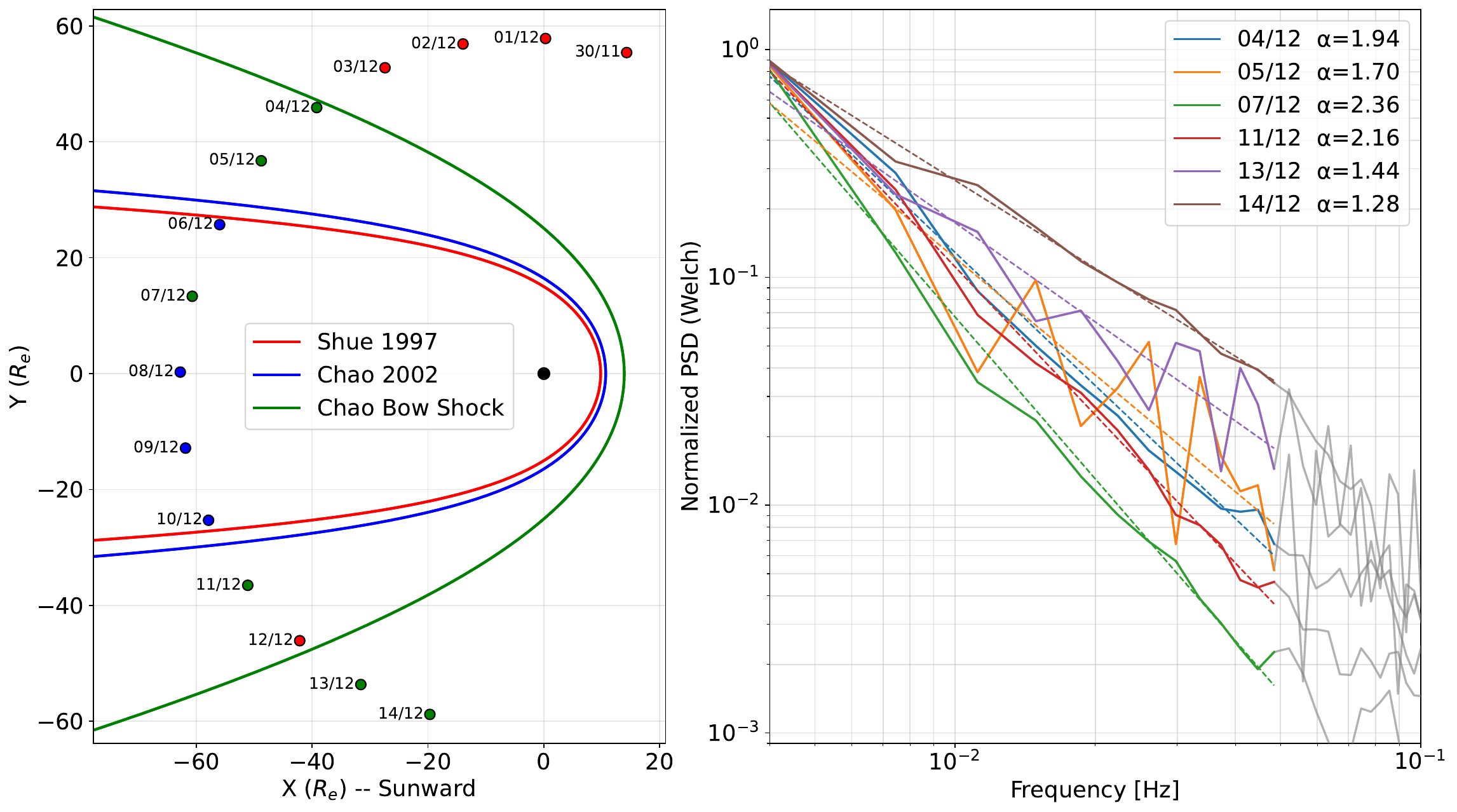}
    \caption{Same as figure \ref{fig:positions_psd} but for DOY 334-348.}
    \label{fig:positions_psd8}
\end{figure}
\section*{Conflict of Interest disclosure}
The authors declare there are no conflicts of interest for this manuscript.
\bibliography{turbulence_refs}

\end{document}